\documentstyle[pra,aps]{revtex}

\draft

\begin{document}
\twocolumn[\hsize\textwidth\columnwidth\hsize\csname@twocolumnfalse%
\endcsname
\title{Gapless Time-Reversal Symmetry Breaking Superconductivity}
\author{A. M. Tikofsky}
\address{Institute for Theoretical Physics, U.C.-Santa Barbara,
Santa Barbara, CA 93106}
\author{D. B. Bailey}
\address{Department of Physics, Stanford University, Stanford, CA 94305}

\date{January 19, 1995}
\maketitle
\begin{abstract}
We consider a layered superconductor with a complex order
parameter whose phase switches sign from one layer to the next.
This system is shown to exhibit gapless superconductivity for
sufficiently large interlayer pairing or
interlayer hopping.  In addition, this
description is consistent with experiments finding signals of time reversal
symmetry breaking in high-temperature superconductors only at the surface
and not in the sample bulk.
\end{abstract}
]

\bigskip
\bigskip

It has been proposed that the many-particle ground state of the
high-temperature superconductors breaks time-reversal symmetry
${\cal T}$\cite{bob-sci,anyons}.
As yet, no clear experimental signatures of ${\cal T}$-violation in the
superconducting bulk have been found \cite{exp-hall,exp-muon}.
Studies of ${\cal T}$-violating ground states
for model magnetic Hamiltonians predict the nontrivially complex order
parameter
$\Delta_{\bf k}=\Delta_0(\cos k_x- \cos k_y+i\epsilon \sin k_x \sin k_y)$
which is known as a $d_{x^2-y^2}+i\epsilon d_{xy}$
($d+id$) order parameter\cite{bob-op}.  This order parameter
does not vanish anywhere on the fermi surface and therefore appears
to be ruled out by
experimental studies of high-temperature
superconductors which indicate the presence of isolated
nodes on the fermi surface\cite{tunneling1,Bonn,heat-capacity}.
A different set of experiments finds
signs of a nodeless ${\cal T}$-violating order parameter
at the surface of the sample.
Evidence of a superconducting gap has been found at isolated
point on the sample surface in STM studies \cite{tunneling2}.
A gap feature of approximately 5 meV is always observed in c-axis tunneling
between YBCO and a conventional superconductor
\cite{tunneling1}.
Anomalous distributions of magnetic flux have been observed
at crystal grain boundaries
in YBa$_2$Cu$_3$O$_7$ \cite{Kirtley}.  This experiment is interpreted
as evidence for
${\cal T}$-violation \cite{Sigrist1}.
In this Letter, we study the properties of a complex order parameter
with a two layer unit cell.  The phase of the order parameter changes
sign from layer to layer.  We show that interlayer pairing and
tunneling render this superconductor gapless at isolated nodes.
Our calculation only describes bulk properties because it relies on the
symmetries of the unit cell.  Because these symmetries
need not be satisfied at sample
boundaries, our model is
consistent not only with gapless superconductivity in the bulk
but also with ${\cal T}$-violating gapped superconductivity at the boundaries.

Let us now investigate the behavior of multilayer
${\cal T}$-violating superconductivity.
Studies of coupled ${\cal T}$-violating systems have found that it is
energetically favorable for the sign of the ${\cal T}$-violation
to be opposite in the two systems\cite{Rojo}.  We therefore choose
the order parameter to be
$\Delta_{\bf k}$ in the first layer and $\Delta_{\bf k}^*$
in the second layer.
We also assume an in-plane spectrum for the non-interacting electrons
of $\epsilon_{\bf k}$, a hopping matrix element between layers of
$t_{\perp{\bf k}}$, and a bilayer order parameter of
$\Delta_{\perp{\bf k}}$.  The standard approach is to define
quasiparticle creation and annihilation operators $a^{\dag}_{{\bf k}\sigma i}$
($i=1,2$ for the first and second planes
respectively, $\sigma=\uparrow,\downarrow$) \cite{Abrikosov}.
The Hamiltonian is then
\begin{eqnarray}
{\cal H}&=&\sum_{{\bf k}\sigma i}
\epsilon_{\bf k}a^{\dag}_{{\bf k}\sigma i} a_{{\bf k}\sigma i}\nonumber \\
&+&\sum_{{\bf k}\sigma} \left[
t_{\perp{\bf k}}a^{\dag}_{{\bf k}\sigma1} a_{{\bf
k}\sigma2}+ t^*_{\perp{\bf k}} a^{\dag}_{{\bf k}\sigma2} a_{{\bf
k}\sigma1}\right]\nonumber \\
&-&\sum_{\bf k}\left[
\Delta_{\perp{\bf k}}(a^{\dag}_{{\bf k}\uparrow1} a^{\dag}_
{{\bf -k}\downarrow2}
+a_{{\bf -k}\downarrow2}a_{{\bf k}\uparrow1}) + \mbox{c.c} \right. \nonumber \\
&\qquad\qquad &+\left.
\Delta_{\bf k}(
a^{\dag}_{{\bf k}\uparrow1} a^{\dag}_{{\bf -k}\downarrow1}
+a_{{\bf -k}\downarrow2}a_{{\bf k}\uparrow2}) + \mbox{c.c} \right].
\end{eqnarray}
This expression can be rewritten using Nambu notation as
\begin{equation}
{\cal K=H-\mu N} = \sum_{\bf k}\Psi^{\dagger}_{\bf k}{\bf Q_k}\ \Psi_{\bf k}
\end{equation}
where
\begin{equation}
{\bf Q_k}\!=\!\left[ \begin{array}{cccc}
\xi_{\bf k} & -\Delta_{\bf k} & t_{\perp{\bf k}} & -\Delta_{\perp{\bf k}} \\
-\Delta_{\bf k}^* & -\xi_{\bf k} & -\Delta^*_{\perp{\bf k}} & -t_{\perp{\bf k}}
\\
t^*_{\perp{\bf k}} & -\Delta_{\perp{\bf k}} & \xi_{\bf k} & -\Delta_{\bf k}^*
\\
-\Delta^*_{\perp{\bf k}} & -t^*_{\perp{\bf k}} & -\Delta_{\bf k} & -\xi_{\bf k}
\end{array} \right]\!,
\Psi_{\bf k}\!=\!\left[ \begin{array}{c}
a_{{\bf k}\uparrow1} \\
a^{\dag}_{{\bf-k}\downarrow1} \\
a_{{\bf k}\uparrow2} \\
a^{\dag}_{{\bf-k}\downarrow2}
\end{array} \right]\!,
\end{equation}
$\xi_{\bf k}=\epsilon_{\bf k}-\mu=-2t(\cos k_x+\cos k_y)-\mu$,
$\Delta_{\bf k}= \Delta_0(\cos k_x-\cos k_y+i\epsilon \sin k_x \sin k_y)$,
and both $\Delta_{\perp{\bf k}}$ and
$t_{\perp{\bf k}}$ are taken to be real.
Writing $\Delta_{\bf k}=\Delta_{R{\bf k}} +i\Delta_{I{\bf k}}$, we
find that ${\bf Q_k}$ has energy eigenvalues $E_{\pm}({\bf k})$ that satisfy
\begin{eqnarray}
\label{spectrum}
\lefteqn{E_{\pm}^2({\bf k})=\xi^2_{\bf k}+ |\Delta_{\bf k}|^2
+\Delta^2_{\perp{\bf k}}+t^2_{\perp{\bf k}}} \nonumber \\
&\pm& 2\sqrt{(\xi_{\bf k}t_{\perp{\bf k}}+\Delta_{\perp{\bf k}}
\Delta_{R{\bf k}})^2+
\Delta^2_{I{\bf k}}(t^2_{\perp{\bf k}}+\Delta^2_{\perp{\bf k}})}.
\end{eqnarray}
The spectrum for $E_{-}({\bf k})$ vanishes when
\begin{equation}
\label{node1}
\xi_{\bf k}=t_{\perp{\bf k}}\alpha_{\bf k}\ ,\qquad
\Delta_{R{\bf k}}=\Delta_{\perp{\bf k}}\alpha_{\bf k},
\end{equation}
and
\begin{equation}
\label{node2}
\alpha^2_{\bf k}=1-
{\Delta^2_{I{\bf k}}\over{t^2_{\perp{\bf k}}+\Delta^2_{\perp{\bf k}}}}\ ,
\end{equation}
which is consistent with
experiments that find a bulk order parameter with a $d$-wave like node
\cite{tunneling1,Bonn,heat-capacity,d-Josephson,d-half}.
If the phase of the order parameter were the
same in neighboring planes, the corresponding energy eigenvalue
$E'_{\pm}({\bf k})$
would satisfy
\begin{equation}
E'^2_{\pm}({\bf k}) = (\xi_{\bf k}\pm t_{\perp{\bf k}})^2
+(\Delta_{I{\bf k}})^2+(\Delta_{R{\bf k}}\mp\Delta_{\perp{\bf k}})^2
\end{equation}
instead of Eq. (\ref{spectrum}) and hence be nodeless except for a special
value of $\Delta_{\perp{\bf k}}$.

In order to have a nodeless order parameter,
we must be able to satisfy Eq. (\ref{node1})-(\ref{node2}).
The constraint that
$\alpha_{\bf k}$ be real requires
\begin{equation}
\label{inequality}
t^2_{\perp{\bf k}}+\Delta^2_{\perp{\bf k}}\ge\Delta^2_{I{\bf k}}\ .
\end{equation}
This intuitive result
tells us that the combined effect of interlayer pairing and interlayer
tunneling has to be sufficiently large in order to overcome the gap
due to the imaginary component of the order parameter.  The existence
of a solution to Eq. (\ref{node1}) also requires
$\Delta_{\bf k}\geq\Delta_{\perp{\bf k}}$.

The zeros of Eq. (\ref{spectrum}) need not
lie on the $45^o$ ($k_x=\pm k_y$) nodal line of the pure $d_{x^2-y^2}$
order parameter.
To first order in $t_\perp/t$ and $\Delta_\perp/\Delta_o$ and in the
$t>>\epsilon\Delta_o$ limit,
the nodal wavevector {\bf k} lies at
\begin{equation}
\label{momentumb}
k_x\approx {\pi\over 2}+\alpha_{\bf k}
\left({t_{\perp{\bf k}}\over{4t}}-
{\Delta_{\perp{\bf k}}\over{2\Delta_o}}\right)\ ,
\end{equation}
\begin{equation}
\label{momentumc}
k_y\approx {\pi\over 2}+\alpha_{\bf k}
\left({t_{\perp{\bf k}}\over{4t}}+
{\Delta_{\perp{\bf k}}\over{2\Delta_o}}\right)\ .
\end{equation}
The node is shifted from the $45^o$ nodal line to an angle
\begin{equation}
\label{theta}
\theta=\arctan(k_y/k_x)\approx
\pi/4+{{\alpha_{\bf k}\Delta_{\perp{\bf k}}}\over{\pi\Delta_o}}
\end{equation}
which is independent of $t_{\perp{\bf k}}$ to lowest order.  In contrast,
the magnitude of {\bf k},
\begin{equation}
\label{momentumf}
|{\bf k}|\approx \sqrt{2}\left ({\pi\over2}+
{{ \alpha_{\bf k} t_{\perp{\bf k}} }\over{4t}}\right )\ ,
\end{equation}
is independent of $\Delta_{\perp{\bf k}}$ to lowest order.
There are now two nodes because
$\alpha_{\bf k}$ may be either positive or negative.
These effects (node splitting and angular shift of 10$^o$)
have been observed in angular resolved
photoemission studies of Bi$_2$Sr$_2$CaCu$_2$O$_8$ \cite{Ding}.
A similar analysis of the momentum shifts has been performed for the
case of a purely real order parameter by P. A. Lee and coworkers
\cite{Kuboki,Ubbens}.
Our expressions are valid in this case with
$\alpha_{\bf k}=\pm 1$ in Eq. (\ref{momentumb})-(\ref{momentumf}).
Therefore, the bulk photoemission experiments cannot distinguish
qualitatively between a purely real and an alternating complex order
parameter.

We may also consider other
alternating complex order parameters.  For example,
we might consider an alternating order parameter that
is purely complex.
Requiring that $\Delta_{R{\bf k}}=0$ is equivalent to imposing
an additional constraint upon Eq. (\ref{node1}).  This equation will only
have a solution in the unlikely case that
either $\Delta_{\perp{\bf k}}=0$ or $\alpha_{\bf k}=0$.
An order parameter with $s+id$ symmetry \cite{s+id} would only
satisfy
Eq. (\ref{node1})-(\ref{node2})
if the $s$ component is smaller than $\Delta_{\perp{\bf k}}$.

The parameters we choose depend on
the material we wish to study.  For example,
neighboring planes in YBCO may be so strongly coupled that they prefer their
order parameters to have the same complex phase.
In that case, the bi-layer
in YBCO may be treated as a single composite
${\cal T}$-violating superconducting layer
coupled to another composite layer with
the opposite sign of ${\cal T}$-violation.
In order to
observe a node, Eq. (\ref{inequality}) must be satisfied
for the coupling of the composite layers.  This description should
be consistent with the spin gap calculations for bilayer
systems \cite{Millis}.
One may also consider 3-layer materials
(which arise in the Bi, Tl, and Hg cuprates) or even the
effect of intermediate
normal layers sandwiched between superconducting layers
with alternate signs of
${\cal T}$-violation.

It is expected that a
${\cal T}$-violating layer
will couple to an electron or neutron spin as if it induced
a local magnetic field.  At long wavelengths, this effect would
be suppressed because of the alternating chirality.
However, a neutron with momentum $q=(0,0,\pi/d)$,
where $d$ is the distance between planes, will couple to
the chirality oscillation.
If the coupling is strong enough, a peak at low energies near this
momentum
should be observable in neutron scattering.  The size of
this feature will depend on the coupling strength, which has
not been accurately calculated.  However,
muon spin rotation experiments, which probe the local magnetic field,
fail to find an effective magnetic field larger than that due to
the Cu nuclei \cite{exp-muon}.  In contrast, similar studies of
UPt$_3$, which is believed to have a $d+id$ order parameter that
does not alternate, saw this effect \cite{upt3}.

Our postulated state can have interesting effects in external magnetic
fields if the energy scales allow these effects to
occur below $H_{c2}$.  The energy gained aligning the
chirality in all of the layers with
a $c$-axis magnetic field may be contrasted
with the energy gained alternating the chirality in
neighboring layers \cite{Rojo}.  If the coupling to the magnetic
field is unusually large, there may be an $H^*<H_{c2}$ above which it becomes
energetically favorable for the layers to violate ${\cal T}$ without
alternating phase.  Hysteretic effects would also be
expected for samples cooled in sufficiently large magnetic fields.

There are three modes made from linear combinations of the
phases of the order parameters $\Delta_{\bf k}$ (the first layer),
$\Delta^*_{\bf k}$ (the second layer), and $\Delta_{\perp{\bf k}}$.
The mode where the phases are added is the
charged Anderson-Bogoliubov mode which is lifted to the plasmon frequency.
The other two linear combinations will have interesting dynamics and,
as recently pointed out by Kuboki et al. \cite{Kuboki},
should be observable as a Raman signal.
Investigation of these modes may provide a
criterion for differentiating between our oscillating
$d+id$ order parameter and the purely real one proposed
by K. Kuboki et al. \cite{phase,future}.
The effect of a complex order parameter on Josephson tunneling
\cite{no-d-josephson,no-d-caxis,beasley-d+id-josephson}
should also be studied.

The ${\cal T}$-violating component of the alternating order parameter
should have a transition temperature $T^*<T_c$ where $T_c$ is the
bulk transition temperature.
The experimental results that we attribute to ${\cal T}$-violation
are therefore expected to disappear for temperatures above $T^*$.
In fact, recent experiments in oxygenated samples of YBCO have found a
second transition at 30K in $T_2$ measurements \cite{Itoh}.

In this Letter we propose that a certain class of ${\cal T}$-violating
superconducting ground states is consistent with numerous experiments.
There remain significant inconsistencies.
Most notably, several experiments have been interpreted as finding
zero Hall conductivity.
In contrast, ${\cal T}$-violation
is consistent with a nonzero value of the Hall conductivity $\sigma_{xy}$.
It is difficult to compare the limits
on the Hall conductivity set by experiment with theoretical
predictions because this quantity has not yet
been accurately calculated \cite{anyons}.
Gauge field fluctuations are expected to reduce the
size of the hall conductivity \cite{anyons,Ubbens,Ubbens2}.
It is also known that
fluctuation effects tend to reduce the quasiparticle gap in single layer
anyon models \cite{gap} and hence the size of the
${\cal T}$-violating order parameter $\Delta_I$.
This effect is expected to be enhanced in
a double layer anyon model \cite{future}.

We have benefited from useful conversations with
M. Beasley, R. B. Laughlin, D. Scalapino, S. C. Zhang,
N. Nagaosa, and M. Sigrist.
This work has been supported by the National Science Foundation
under grants
PHY89-04035 (A.M.T.) and DMR88-16217 (D.B.B.).  D. B. Bailey
gratefully acknowledges the support of an NSF fellowship.

\end{document}